\documentclass[preprint,aps,showpacs,showkeys]{revtex4}
\usepackage{amsmath}

\begin{document}

\preprint{}

\title{Thermoeconomic analysis of an irreversible Stirling heat pump cycle}

\author{Umberto Lucia}
\email[E-mail address: ]{umberto.lucia@istruzione.it}
\affiliation{I.T.I.S. "A. Volta", Spalto Marengo 42, 15100 Alessandria, Italy}

\author{Gianpiero Gervino}
\email[E-mail address: ]{gervino@to.infn.it}
\affiliation{Dipartimento di Fisica Sperimentale, 
	Universit\`a di Torino, Via P. Giuria 1, 10125 Torino, Italy}

\date{\today}

\begin{abstract}
In this paper an analysis of the Stirling cycle in thermoeconomic
terms is developed using the entropy generation. In the thermoeconomic optimization of an
irreversible Stirling heat pump cycle the F function has been
introduced to evaluate the optimum for the higher and lower sources
temperature ratio in the cycle: this ratio represents the value which
optimizes the cycle itself.  The variation of the function F is proportional to
the variation of the entropy generation, the maxima and minima of F has been evaluated in
a previous paper without giving the physical foundation of the method.  We
investigate the groundwork of this approach: to study the
upper and lower limits of F function allows to determine the cycle stability and the
optimization conditions. The optimization consists in the best $COP$ at
the least cost. The principle
of maximum variation for the entropy generation becomes the analytic
foundation of the optimization method in the thermoeconomic analysis
for an irreversible Stirling heat pump cycle.
\end{abstract}
\maketitle{}

\section{Introduction}
\label{sec:intro}

The Stirling cycle is an important model of refrigeration systems and
the recent developments in design were proposed after the new concept
of finite time thermodynamics came into existence \cite{1,2,3}. Blanchard
applied the Lagrange multiplier method to find out the $COP$ of an
endoreversible Carnot heat pump operating at the minimum power input
for a given heating load \cite{4}. Several papers heve been devoted to propose mathematical functions to optimize thermodynamics cycles starting from different initial conditions and focusing on the total cost and efficiency.  The definition of optimization that we adopt in this paper is the best $COP$ at the least cost. The performance of the
different heat engine and refrigeration systems were investigated
using the concept of finite time thermodynamics, of the ecological
approach and the thermoeconomic analysis \cite{1,5,6,7,8}. On the other hand,
the key-role of entropy generation maximum has been recently demonstrated in thermodynamics analysis of the irreversible processes and it has been shown that it represents  a new criterion for determining the conditions for stability \cite{9,10,11,12,13,14}. In this paper we show that this criterion represents the thermodynamic foundation for some recent results obtained in thermoeconomic analysis of the Stirling heat pump cycle. We start from studying the time evolution of an open system and we take as working hypothesis that it evolves in the optimization of the irreversibilitiy due to entropy generation.   F function is well suited to show the system evolution related to the irreversibility due to entropy generation, our aim is to propose an approach based upon the natural behaviour of the thermodynamics/thermoeconomic system as a groundwork for the optimization analysis.  The evolution of an open system is considered natural when it moves in order to get the optimization of the entropy generation.

\section{The thermodynamic analysis}
\label{sec:2}

The working substance of the Stirling cycle may be a gas, a magnetic
material, etc., and for different working fluids the performance
of the cycle are quite different. The Stirling cycle
with an ideal gas consists of two isothermal and two isochoric
processes. It approximates the expansion stroke of the real cycle by an
isothermal process to whom heat is added to reach the temperature T$_{c}$ from a heat source of finite capacity whose temperature varies
from T$_{L1}$ to T$_{L2}$. The heat addition to the working fluid is
thought as an isochoric process: heat is going towards the
heat sink of finite heat capacity that gets a temperature variation from
T$_{H1}$ to T$_{H2}$. The heat rejection from the working fluid to the
regenerator is modelled as an isochoric process which completes the
cycle itself. Let Q$_c$ and Q$_h$ be the amount of heat absorbed from
the sources at the temperature T$_c$ and T$_h$ respectively, during
the two isothermal processes \cite{1}:

\begin{equation}
  \label{eq:s1}
  Q_h = C_H\epsilon_H(T_h-T_{H1})t_h
\end{equation}
\begin{equation}
  \label{eq:s2}
  Q_c = C_L\epsilon_L(T_{L1}-T_{c})t_L
\end{equation}

where $C_H$ is the heat capacitance rate of the sink reservoir, $C_L$
is the heat capacitance rate of the source reservoir, $t_H$ is the heat
rejection time, $t_L$ is the heat addition time, $\epsilon_H$ is the
effectiveness of the heat exchangers for the hot-side and $\epsilon_L$
is the effectiveness of the heat exchangers for the cold-side. These
cycles do not possess the condition of perfect regeneration, hence it is
assumed that the loss per cycle, $\Delta Q_R$, is proportional to the
temperature difference of the two isothermal processes as
follows \cite{1,15,16,17,18,19}:
\begin{equation}
  \label{eq:s3}
  \Delta Q_R = nc_f(1-\epsilon_R)(T_h-T_c)
\end{equation}

where $c_f$ is the molar heat capacity of the working fluid and $n$ is the
number of moles. The Gouy-Stodola theorem \cite{20} states that the thermodynamic work burnt in the irreversibility due to the entropy generation is equal to the product between the lowest source temperature and the entropy generation, i.e. total entropy is equal to the isolated system entropy plus the irreversibility due to entropy generation.  Considering the Gouy-Stodola theorem  and the
definition of the entropy due to irreversibility $\Delta S_{irr}$ \cite{13}, the last one can be written as:

\begin{equation}
  \label{eq:s4}
  \Delta S_{irr} = \frac{\Delta Q_R}{T_c}=nc_f(1-\epsilon_R)\frac{(T_h-T_c)}{T_c} = nc_f(1-\epsilon_R)(x-1)
\end{equation}

with \(x=T_h/T_c\). The theorem of maximum entropy generation states that
the entropy generation is maximum at stationary state \cite{9}. This theorem
allows a new approach  to irreversible processes as it is proved in a lot of
different applications in hydrodynamics \cite{10}, engineering thermodynamics
\cite{11}, rational thermodynamics   \cite{12} and biophysics \cite{13,14}. Hence
applying it here, we argue that equation \ref{eq:s4} must be a maximum in the
thermodynamics stability: this equation described the natural behavior of the thermodynamics system.

\section{The thermodynamics foundation of the thermoeconomic analysis}
\label{sec:3}

The objective function F of the thermoeconomic optimization recently
proposed is \cite{1,21,22}:

\begin{equation}
  \label{eq:s5}
  F = \frac{\dot Q_H}{C_i + C_e}
\end{equation}

with $\dot Q_H =$ heating power, $C_i$ and $C_e$ refer to annual investment and energy
consumption costs, and are defined as:
\begin{equation}
  \label{eq:s6}
 C_i = a(A_H+ A_L+ A_R)+b\frac{Q_h-Q_c}{t_{cycle}}
\end{equation}

\begin{equation}
  \label{eq:s7} 
  C_e=b\frac{Q_h-Q_c}{t_{cycle}}
\end{equation}


where $a$  is a constant directly proportional to the investment cost of
the heat exchanger and is equal to the capital recovery factor multiplied by
the investment cost per unit heat exchanger area.  $A_H+ A_L+ A_R$ is the
heat exchanger total area, with $A_H$ the heating area, $A_L$ the heat
source area and $A_R$ the regenerative area. $b$ is the capital recovery
factor multiplied by the investment cost per unit power input and $t_{cycle}$ is
defined as:

\begin{equation}
  \label{eq:s8} 
  t_{cycle} = t_H+t_L+t_R
\end{equation}

with

\begin{equation}
  \label{eq:s9} 
t_R=2\alpha(T_h - T_c)=2\alpha T_c(x-1)	
\end{equation}

where $\alpha$ is a constant that depends upon the kind of working fluid used
in the cycle, and shows that the working time of the regenerator (a sort of recovering time towards the initial conditions in thermoeconomics) is proportional to the difference of temperature. In the thermoeconomic analysis of an irreversible
Stirling heat pump cycle the function F has been used to evaluate the
ratio of the higher and lower source temperature in order to reach the
optimization of the cycle itself. The common solution, based upon the
application of the variation method, consists in evaluating the maxima of 
F function, solving the equation $\delta F = 0$ \cite{1} applying the variational method.

Now, from \ref{eq:s6}  and \ref{eq:s7} the \ref{eq:s5} becomes:

\begin{equation}
  \label{eq:s10}
  \begin{split}
    F&=\cfrac{\dot Q_H}{a(A_H+ A_L+ A_R)+(b+b')\cfrac{Q_h-Q_c}{t_{cycle} }}
\end{split}
\end{equation}	

Starting from the relations \ref{eq:s4}, \ref{eq:s5}, \ref{eq:s8}-\ref{eq:s10}, we can argue that the
objective function F of the thermoeconomic optimization is related to
the entropy generation as follows:

\begin{equation}
  \label{eq:11}
  \begin{split}
    F&=\cfrac{\dot Q_H}{a(A_H+ A_L+
      A_R)+(b+b')\cfrac{Q_h-Q_c}{t_H+t_L+\cfrac{2\alpha T_c}{nc_f(1-\epsilon_R)}\Delta
        S_{irr}}} 
\end{split}
\end{equation}	

which can be easily written after few algebric operations:
\begin{equation}
  \label{eq:12}
  \begin{split}
    F =
    &= \frac{\Gamma_1+\Gamma_2\Delta S_{irr}}{\Gamma_3+\Gamma_4\Delta S_{irr}}
\end{split}
\end{equation}

\begin{equation*}
  \text{with}\qquad
  \begin{cases}
    \Gamma_1 =& \dot Q_H(t_H+t_L)\\
    \Gamma_2  =& \frac{2\alpha T_c\dot Q_H}{nc_f(1-\epsilon_R)}\\
    \Gamma_3  =& a(A_H+ A_L+ A_R)(t_H+t_L)+(b-b')(Q_h-Q_c)\\
    \Gamma_4  =& \frac{2\alpha a(A_H+A_L+A_R)T_c}{nc_f(1-\epsilon_R}
  \end{cases}
\end{equation*}

From equation \ref{eq:12} we can argue that the variation of the function F is
proportional to the variation of the entropy generation:

\begin{equation}
  \label{eq:s14}
 \delta F=\frac{\Gamma_2\Gamma_3-\Gamma_1\Gamma_4}{\Gamma_3+\Gamma_4\Delta S_{irr}}\delta(\Delta S_{irr})
\end{equation}

with

\begin{equation}
  \label{eq:s15}
\Gamma_2\Gamma_3 \neq \Gamma_1\Gamma_4
\end{equation}

Then it follows that 

\begin{equation}
  \label{eq:s16}
  \delta(\Delta S_{irr})=0\Rightarrow\delta F = 0
\end{equation}

In this way it has been
stressed the relation between the economic analysis and the
thermodynamics.  In the economical analysis the function F was
introduced in several papers: we need to know its upper and lower limits to fulfill the basic conditions of optimization, but no physical explanation has
been up to now given about this method.  Here we prove that the
limits of the F function are directly correlated to the 
entropy generation in the state of stability and related to the
optimization of the cycle.  Hence the optimization, which
consists in the best COP related to the least cost, can be obtained
in the conditions of natural stability for the open systems.  The evolution of an open system is defined natural when it moves to get the optimization of  entropy.   The advantages of this method consist in exploiting the natural dynamics of the system in order to reach, following its natural behaviour, the optimum by the shortest way (i.e. the lower cost).

\section{Conclusions}

The thermodynamic and thermoeconomic analysis of the optimization of
an irreversible Stirling heat pump cycle is presented in relation with
its thermodynamic foundation. We proved that the principle of maximum
variation for the irreversible entropy is the analytic foundation for
the optimization method recently introduced in the thermoeconomic
analysis for an irreversible Stirling heat pump cycle. Of course it
represents not only an analytical and mathematical groundwork, but
also the physical and thermodynamic foundation for the method itself,
as a consequence of the physical meaning of the principle of maximum
entropy variation in thermodynamics \cite{9,10,11,12,13,14}.
The optimization method is a useful tool to design thermodynamics systems
characterized by lower working costs. The principle of maximum variation allows
a deeper thermoeconomic analysis focused on the stability conditions.

\subsection*{Nomenclature}

\begin{tabular}{lll}

$a$ & capital recovery factor times cost& \\
 &  per unit heat 0 area& \\
$A$ & area & $[m^2]$ \\
$b$ & capital recovery factor times investment cost &         \\ 
 &  per unit power input     &         \\ 
$c$ & molar heat capacity     &  $[J mole^{-1} k^{-1}]$       \\ 
$C$ & heat capacitance rate     & $[kW K^{-1}$        \\ 
$COP$ & Coefficient of Performance     &         \\ 
$n$ & number of moles     &   $[mole]$      \\ 
$Q$ & heat     &  $[J]$       \\ 
$S$ &  entropy    &  $[J K^{-1}]$       \\ 
$t$ &  time    &   $[s]$      \\ 
$T$ &  temperature    & $[T]$        \\ 
$x$ & $\frac{T_h}{T_c}   $     &         \\ 
    &      &         \\ 
\textit{Greek letters} &      &         \\ 
$\epsilon$ & effectiveness     &         \\ 
$\delta$ &  differential 1 $\sum_{i} \frac{\partial}{\partial{z_{i}}} dz_{i} $   &         \\ 
$\Delta$ &  finite variation    &         \\ 
    &      &         \\ 
\textit{Subscripts} &      &         \\ 
$f$ &  fluid    &         \\ 
$h$ &  sink side    &         \\ 
$H$ &  heating    &         \\ 
$irr$ & irreversible which is related     &         \\ 
 &  to the entropy generation     &         \\ 
$L$ & heat source     &         \\ 

\end{tabular}

\newpage{}

\end{document}